\documentclass[preprint]{aastex}

\usepackage{amsmath}
\usepackage{natbib}
\usepackage{url}
\usepackage[english]{babel}
\usepackage{graphicx}
\usepackage{txfonts}
\usepackage{subfigure}
\usepackage{aas_macros}
\newcommand{\ttt}{\times 10^}

\shorttitle{Particle acceleration by filamentation instabilities}
\shortauthors{Burkart et al.}

\begin{document}

\title{The influence of the mass-ratio on the acceleration of particles by filamentation instabilities}

\author{Thomas Burkart}
\author{Oliver Elbracht}
\author{Urs Ganse}
\author{Felix Spanier}
 \email{fspanier@astro.uni-wuerzburg.de}
 \affil{Lehrstuhl f\"ur Astronomie, Universit\"at W\"urzburg,
  	Am Hubland, D-97074 W\"urzburg\\}
\date{\today}

\begin{abstract}

  Almost all sources of high energy particles and photons are
  associated with jet phenomena. Prominent sources of such highly relativistic outflows are pulsar winds, Active
  Galactic Nuclei and Gamma-Ray Bursts. The current understanding of
  these jets assumes diluted plasmas which are best described as
  kinetic phenomena. In this kinetic description particle acceleration to ultra-relativistic speeds
  can occur in completely unmagnetized and neutral plasmas through insetting effects of
  instabilities. Even though the morphology and nature of  particle spectra are
  understood to a certain extent, the composition of the jets is not 
  known yet. While Poynting-flux dominated jets (e.g. occuring
  in pulsar winds) are certainly composed of electron-positron plasmas, the
  understanding of the governing physics in AGN jets is mostly unclear.

  In this article we investigate how the constituting elements of an electron-positron-proton plasmas behave differently under the variation of the fundamental mass-ratio $m_{p} / m_{e}$.

  We studied initially unmagnetized counterstreaming plasmas using fully
  relativistic three-dimensional particle-in-cell simulations to
  investigate the influence of the mass-ratio on particle acceleration
  and magnetic field generation in electron-positron-proton plasmas. We
  covered a range of mass-ratios $m_{p} / m_{e}$ between 1 and 100 with
  a particle number composition of $n_{p^+} / n_{e^+}$ of 1 in one stream, therfore called the pair-proton stream. Protons are injected
  in the other one, therfore from now on called proton stream, whereas electrons are present in both to guarantee charge neutrality in the simulation box.
  We find that with increasing proton mass the instability takes longer to
  develop and for mass-ratios $> 20$ the particles seem to be
  accelerated in two phases which can be accounted to the individual
  instabilities of the different species. This means that for high mass
  ratios the coupling between electrons/positrons and the heavier protons, which occurs in low mass-ratios, disappears.
\end{abstract}


\maketitle


\section{\label{introduction}Introduction}
Radiation observed from astrophysical systems like GRBs or AGN usually possesses a nonthermal emission spectrum. This is believed to arise from particle acceleration in the vicinity of relativistic shocks or within the counterstreaming plasma itself.

In recent PiC simulations it has been shown that most particle acceleration occurs within the jet \citep{2003ApJ...595..555N, 2005ApJ...622..927N, 2006ApJ...642.1267N, 2008AIPC.1085..589N, 2009arXiv0901.4058N, 2008ApJ...674..378C, 2008ApJ...673L..39S, 2008ApJ...675..586D, 2004ApJ...608L..13F, 2004ApJ...617L.107H, 2005ApJ...623L..89H, 2009ApJ...695L.189M, 2007ApJ...671.1877R, 2003ApJ...596L.121S} and is mostly caused by plasma instabilities like the Weibel \citep{1959PhRvL...2...83W} or twostream instability \citep{1958PhRvL...1....8B}. Both instabilities create current filaments with surrounding magnetic fields and are therefore a plausible source for particle acceleration and the generation of observed long-lasting magnetic fields.
Particle acceleration can also occur along with shocks where first order Fermi acceleration \citep{1949PhRv...75.1169F} is assumed to be the relevant process, which was shown by kinetic simulations only recently \citep{2008ApJ...682L...5S}.

In the present work we focus on the main properties of the plasma instabilities and describe the influence of the fundamental mass-ratio $m_p/m_e$ in mixed electron-positron-proton compositions by means of relativistic three-dimensional simulations of counterstreaming plasmas.

The paper is organized as follows: In chapter \ref{description}  the underlying code is described briefly, in section \ref{setup} we illustrate the setup of the performed simulations. In chapter \ref{results} we present the results of our simulations which we discuss and draw some conclusion in section \ref{discussion}.

\section{\label{description}Description of the code}
Particle-in-Cell simulations are an essential tool in understanding relativistic collisionless plasma physics. Therefore we developed a three-dimensional fully relativistic MPI-parallelised PiC code called ACRONYM (\textbf{A}nother \textbf{C}ode for moving \textbf{R}elativistic \textbf{O}bjects, \textbf{N}ow with \textbf{Y}ee lattice and \textbf{M}acroparticles). Maxwells equations are evolved in time by employing a second-order leapfrog scheme (see e.g.~\citealt{Taflove}). The particles affect the electromagnetic fields through charge currents which are deposited on the grid by using a second-order Triangular Shaped Cloud (TSC) scheme (see e.g.~\citealt{1988csup.book.....H}) adopted from \citet{2001CoPhC.135..144E}. The particles are moved via a second-order force interpolation within the Boris push \citep{Boris1970}. In order to guarantee that the divergence of the magnetic field remains close to zero, the electric and magnetic fields are stored in the form of a staggered grid, the so-called Yee-lattice \citep{1966ITAP...14..302Y}. With this setup the code is second-order both in space and time.

Extensive tests of the code have been successfully completed from which we conclude that the total relative error in energy conservation is less that $3\ttt{-5}$ and the divergence of the magnetic field stays below a value $\left| \nabla \vec{B}/B \right| < 10^{-12}/\lambda_D$ in the simulated space for all times.

\section{\label{setup}Simulation setup}

In the simulations presented here we use two counterstreaming plasma
populations, one representing the background medium consisting of $6
e^-$ and $6 p^+$ per cell (proton stream) and the other incorporating the jet
containing $4 e^-$, $2 e^+$ and $2 p^+$ per cell (pair-proton stream), from which we find
the background density ratio $n_{jet}/n_{bg} = 2/3$ and the ratio
$n_{p^+}/n_{e^+} = 1$ in the pair-proton stream. In the lab frame (the rest frame of
the simulation box) the two streams are
counterstreaming along the z-direction with a Lorentz factor $\gamma =
10$ ($\beta = v/c = 0.995$) each, the electron distribution has a
thermal velocity of $v_{th,e} = 0.1c$ in every direction in the
restframe of the moving medium, the thermal velocity of the protons is
$v_{th,p} = 0.1c \cdot (m_p/m_e)^{-1}$. This setup resembles
situations as they are believed to exist in jets running into the
interstellar or intergalactic medium.

Three-dimensional simulations with five different compositions of counterstreaming plasmas using $128 \times 128 \times 512$ cells with a total of 167 million particles (20 particles per cell) and mass-ratios $m_{p}/m_{e}$ between 1 and 100 have been performed. In addition to that another simulation with a mass-ratio of 100 with twice the number of cells in each of the perpendicular directions (and therefore four times more particles) has been performed in order to show the influence of the periodic boundary conditions. As pointed out by \citet{2008PPCF...50l4034F}, 20 particles per unit cell (on average) in combination with quadratic particle interpolation are sufficient to eliminate most of the numerical noise. Nevertheless, we have performed simulations up to twice the numbers of particles per cell and no significant changes were observed.

Periodic boundary conditions have been applied in all three dimensions. Due to the quick development of the selfconsistent electromagnetic fields it is redundant to solve Poissons equation at the initial time and the fields can be initialized with zero without loss of generality.

The cell size is set to be equal to the Debye-length of the plasma, $\Delta x = \lambda_D = (\mathrm{k_B} T / 4 \pi n_e e^2)^{1/2}$, and the timestep is restricted by the CFL-criterion, $c \cdot \Delta t < \Delta x / \sqrt{3}$. This results in a $\Delta t$ between $0.035 \omega_p^{-1}$ and $0.050 \omega_p^{-1}$ (normalized to the plasma frequency $\omega_{p} = (4 \pi e^2 n / m_e)^{1/2}$) and the cellsize $\Delta x$ ranges from $2\ttt4\ \mathrm{cm}$ to $3\ttt4\ \mathrm{cm}$ depending on the different mass-ratios employed. The simulations were evolved for 2500 to 4500 timesteps to roughly 120 to 220 $\omega_p^{-1}$ (the exact numbers for each simulation can be found in Table \ref{simusetup}). The characteristic scales of interest in a counterstreaming electron-positron-proton plasma are of the order of several proton skin depths, $c/\omega_{pi} = (\gamma m_p c^2 / 4 \pi e^2 n)^{1/2}$, for a plasma with the density $n$ and the average proton energy $\gamma m_p c^2$.

\begin{table}
\begin{tabular}{|r|r|r|r|r|}
  \hline $m_p/m_e$ & $\Delta t [\omega_p^{-1}]$ & $\Delta x [\mathrm{cm}]$ & number of cells & number of ion skin depths \\
  \hline 1.0   & 0.0498 & 23225 & $128 \times 128 \times 512$ & $120 \times 120 \times 480$ \\
  \hline 5.0   & 0.0385 & 23107 & $128 \times 128 \times 512$ & $57 \times 57 \times 228$ \\ 
  \hline 20.0  & 0.0360 & 21614 & $128 \times 128 \times 512$ & $28 \times 28 \times 112$ \\
  \hline 42.8  & 0.0355 & 21338 & $128 \times 128 \times 512$ & $20 \times 20 \times 80$ \\
  \hline 100.0 & 0.0408 & 21198 & $128 \times 128 \times 512$ & $13 \times 13 \times 52$ \\
  \hline 100.0 & 0.0574 & 29830 & $256 \times 256 \times 512$ & $36 \times 36 \times 72$ \\
  \hline
  \end{tabular}
  \caption{Setup of the simulations performed}
\label{simusetup}
\end{table}


\section{\label{results}Simulation results}

The results of our simulations can be divided into two main findings: (1) new insights into the evolutionary behavior of twostream instabilities in multi-component plasma and (2) the change in the distribution function of the particles, in particular the acceleration caused by the instability.

Due to the huge amount of data, simulation results have been written every tenth step for the fields (electric and magnetic fields, currents) and every hundredth step for particle data.

\subsection{\label{results_fields}Analysis of the fields and currents}
In this section we analyze the evolutionary behaviour of the electric and magnetic fields and the currents in the simulations conducted. From previous simulations of filamentation instabilities in pair plasmas (see e.g.~\citealt{2003ApJ...596L.121S}) it is well known how magnetic and electric fields evolve. We compare the behaviour of plasmas with different mass-ratios. The most significant quantity in this context is the transverse magnetic field energy averaged over the entire computational domain $B_{\perp}^2 = (B_x^2 + B_y^2)$, since strong magnetic fields are essential to create and maintain the flux tubes observed in kinetic instabilities, furthermore the point in time the instability peak occurs and also the existence of a second peak, respectively.

\begin{figure}[ht]
  \rotatebox{270}{ \includegraphics[width=0.65\textwidth]{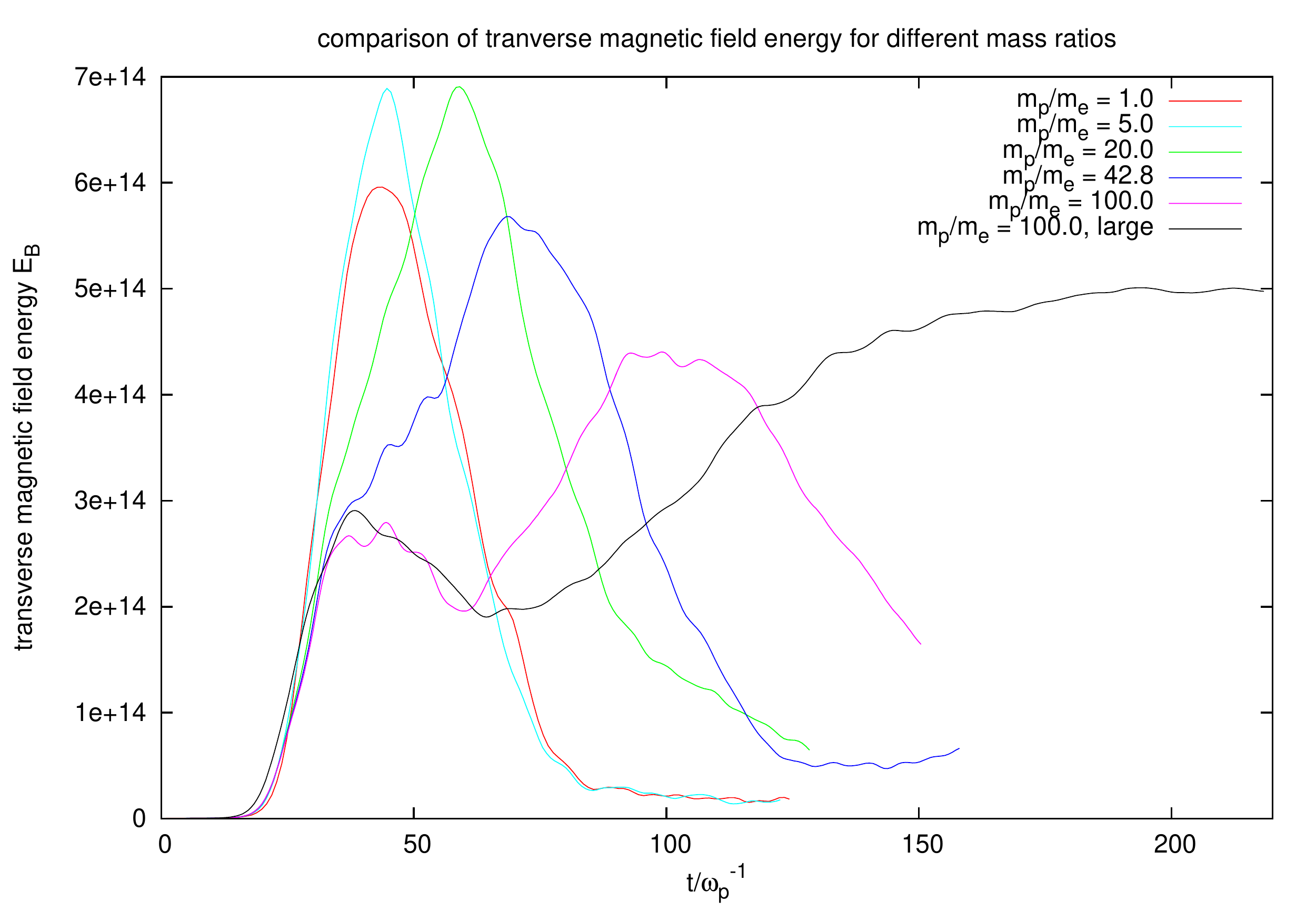} }
  \caption{\label{B_vergleich}Comparison of the time evolution of the transverse magnetic field energy for different mass-ratios. With increasing mass-ratio the instability takes longer to develop. For the two highest mass-ratios one notices that the instability develops in two phases.}
\end{figure}

In Fig.~\ref{B_vergleich} we therefore compare the time evolution of the transverse magnetic field energy $B_{\perp}$ computed in the lab frame as function of different mass-ratios $m_p/m_e$.

It is evident that the maximum value of the transverse magnetic field energy reached in the different simulations are comparable, even though it has to be noted that for the non-pair plasma the maximum energy can only be found in the second peak. The development of the second peak shows a nicely observable dependence on the fundamental mass ratio: For the lower mass-ratios no two peak structure can be seen, while with increasing mass-ratio a clear distinction can be made.

When looking at the time until the instability fully develops, one can see that for higher mass-ratios $m_p/m_e$ it takes longer to reach the peak value for the magnetic field. If one compares simulations with mass-ratios of 1 and 100 one can explain what is happening in the counterstreaming plasma: First an electron-positron instability develops almost simultaneously for both mass-ratios (single peak for mass-ratio 1 and first peak for mass-ratio 100). If a third and heavier species exists another peak will be apparent at later times (which can be seen in Fig.~\ref{B_vergleich}). This behavior is not observable for medium mass-ratio simulations since both peaks overlap and can not be distinguished anymore.

The existence of two instabilities in the plasma has important impact on the amplitude and duration of the instability: Clearly the instability lasts longer for the high mass-ratio simulations, since in this case the heavy protons are accelerated slower compared to the lighter electrons/positrons but are able to stabilize the flux tubes for a longer period of time. Another result to note is that the maximum amplitude decreases with increasing mass-ratio. This effect can be attributed to the lower number of particles constituting each instability.

\begin{figure}[ht]
  \rotatebox{270}{ \includegraphics[width=0.9\textwidth]{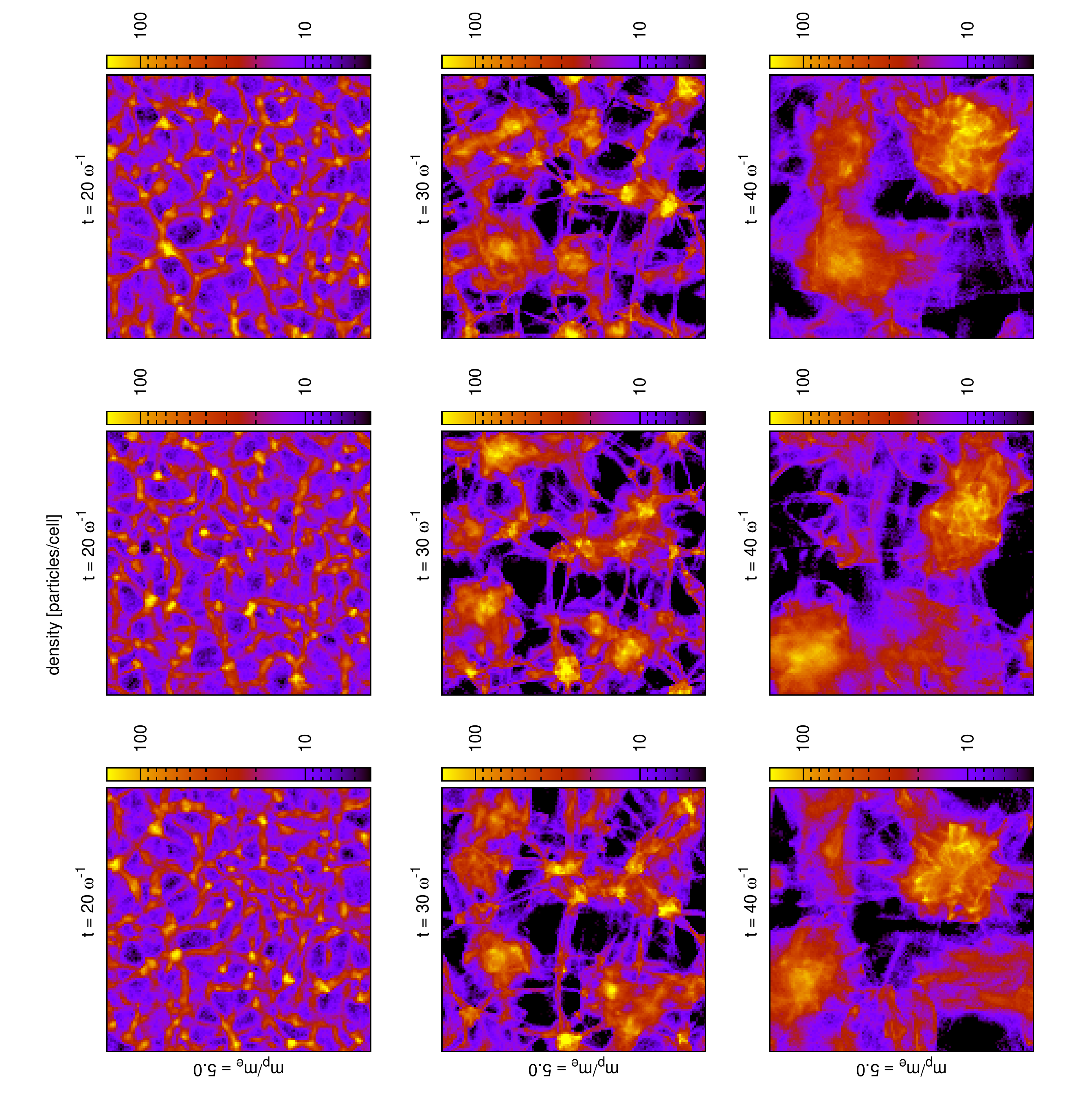} }
  \caption{\label{9_slices_5}Development and merging of the flux tubes for the simulation with $m_p/m_e = 5.0$ and a resolution of $128 \times 128 \times 512$ cells. The colors pertain to the particle density (in particles per cell) shown at three different locations along the the direction of streaming (at 100, 300 and 500 $\Delta x$). One can see the fluxtubes developing and merging over time. Both fluxtubes have about the same strength, which shows that the instabilities of the electrons/positrons and the protons are still srongly coupled for this mass-ratio.}
\end{figure}

\begin{figure}[ht]
  \rotatebox{270}{ \includegraphics[width=0.9\textwidth]{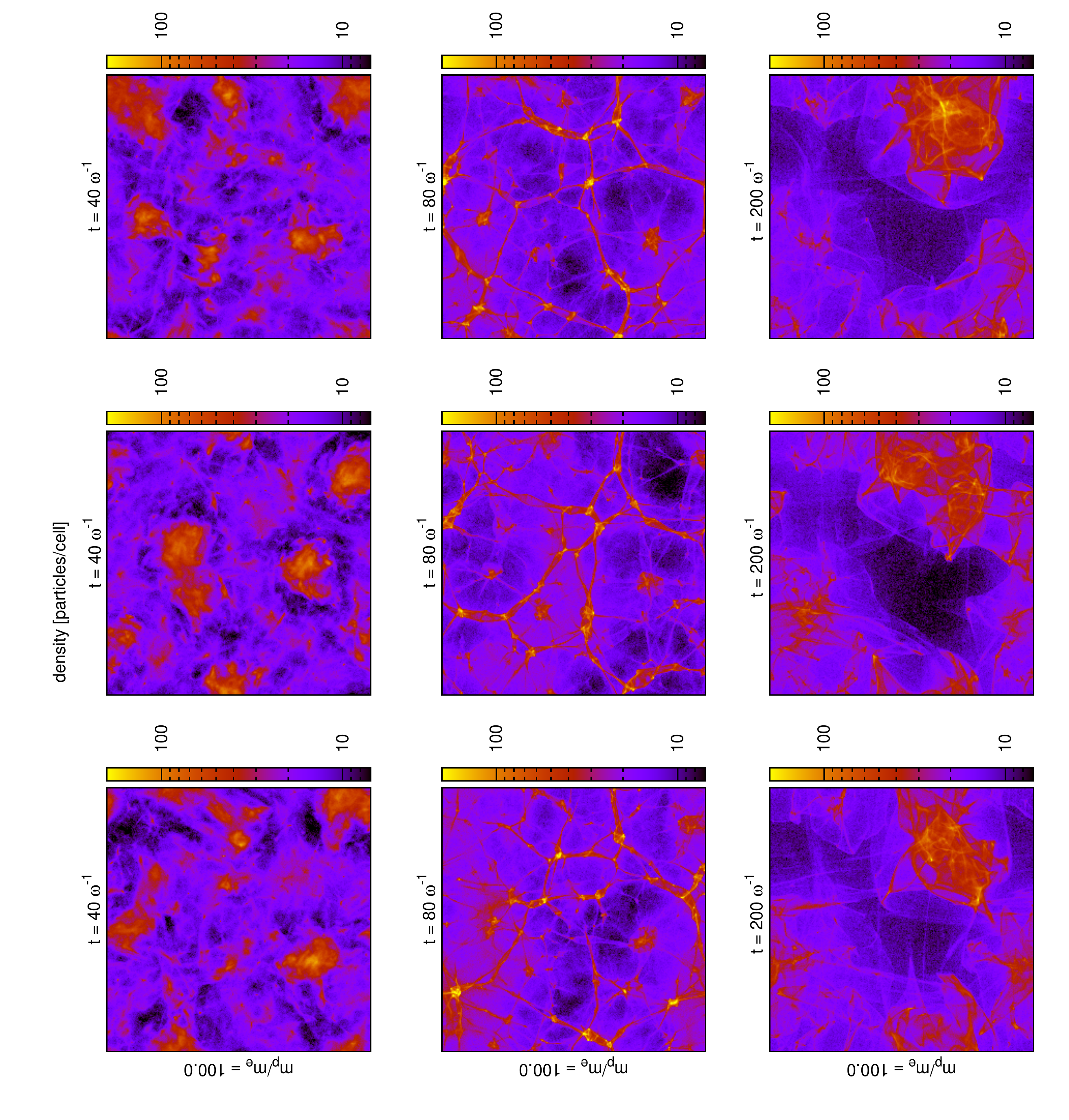} }
  \caption{\label{9_slices_100}Development and merging of the flux tubes for the simulation with $m_p/m_e = 100.0$ and a resolution of $256 \times 256 \times 512$ cells. The colors pertain to the particle density (in particles per cell) shown at three different locations along the the direction of streaming (at 100, 300 and 500 $\Delta x$). In the upper three pictures it is to be seen that first the flux tubes develop, later they nearly vanish (which corresponds to the dip in the magnetic field energy in Fig. \ref{B_vergleich} in the black curve at $80 \omega_p^{-1}$) and then the flux tube in the lower right corner grows stronger (lower set of pictures). This can be attributed to the independent instabilities at high mass-ratios.}
\end{figure}

To inspect the effect of employed mass-ratios on the temporal evolution of instabilities, we looked at the nature of the flux tubes more closely. The flux tubes in the simulations with mass-ratios of $m_p/m_e = 5.0$ and $m_p/m_e = 100.0$ are illustrated in Fig.s~\ref{9_slices_5} and \ref{9_slices_100}, respectively, which show the particle number density in particles per cell at three different locations perpendicular to the direction of streaming. The pictures in Fig.~\ref{9_slices_5} are chosen such that the uppermost row shows the onset of the instability and the lowest pictures are roughly taken at the time the maximum of the instability occurs (cf.~Fig.~\ref{B_vergleich}). In Fig.~\ref{9_slices_100} the upper row of slices is taken at the moment the electron/positron instability peaks, the second set of pictures show the time between the two instabilities (compare with the black curve in Fig. \ref{B_vergleich} at $80 \omega_p^{-1}$) and the last set shows the point in time when the proton instability reaches its peak. 
Both simulations show the archetypical behavior of filamentation instabilities: Flux tubes develop, which in turn merge until only two flux tubes survive. But for the high mass-ratio simulations this whole process happens twice. In an early stage (which resembles the first peak of the instability in Fig.~\ref{B_vergleich}) flux tubes arise. In a later stage (second peak) flux tubes of different strengths exist, one of them is more pronounced. The explanation is that during the first stage of the instability the flux tubes are carrying more of the lighter particles. The second stage is then associated with a flux tube of heavier particles which takes longer to develop, but is also able to exist for a much longer timespan.

The combination of Fig.s~\ref{B_vergleich}, \ref{9_slices_5} and \ref{9_slices_100} suggests that the instability is evolving in two phases. In the first phase light particles are accelerated and in the second phase the heavier particles are also involved in the instability. For mass-ratios of $m_p/m_e \lesssim 20.0$ the instabilities and therefore particle acceleration of the two species cannot be separated. There are two possible reasons for this: Either the coupling of the two species is still strong enough to co-accelerate the heavier particles or the time scales of the two instabilities are still matching rather well. A strong argument for the second option is the fact that the instability time scale for the heavy species increases less than linear with the mass-ratio for a certain size of the computational domain.

Considering only mass-ratios $m_p/m_e > 20.0$ the instabilities are clearly separated which is also apparent in the double-hump structure in the energy-diagrams of Fig.~\ref{B_vergleich} and in Fig.~\ref{9_slices_100} illustrating the development of the flux tubes in two phases.

In the larger simulation (twice the size in the perpendicular directions) it takes even longer for the proton instability to develop because the flux tubes have more space to develop and therefore it takes longer for them to merge until only two are left. This is also the reason for the slightly different slopes in the two simulations with mass-ratio 100. The maxima and minima of the magnetic energy occur around the point in time, when the current density in the direction of streaming averaged over the whole computational domain changes its sign. When (in the larger simulation) the proton instability kicks in (at around $65\, \omega_p^{-1}$) the flux tubes are not yet fully merged down to two and therefore the proton instability does not grow with the same rate as in the smaller simulation. The resulting flux tubes around the maximum of the second instability still resemble the two flux tube regime as seen in the smaller simulations.

\subsection{\label{results_mass}Particle distribution}

As described in chapter \ref{setup} all simulations were initialized with thermal particle distributions (width of the thermal distribution $0.1 c$ and $0.1 c \cdot (m_p/m_e)^{-1}$ for electrons and protons, respectively) which are then boosted with a Lorentz factor of $\gamma = 10$ in either direction. While some particles gain a lot of energy during the simulation in total, the shape of the particle distribution is also changing. To analyze and quantify the change of the particle energy distributions we utilize two distinct types of graphs: (1) a two-dimensional plot in the lab frame relating the absolute value of the momentum parallel ($v_{||}/c \cdot \gamma$ with $v_{||} = \left| v_z \right|$) and perpendicular ($v_{\perp}/c \cdot \gamma$ with $v_{\perp} = (v_x^2 + v_y^2)^{\frac{1}{2} }$) to the initial streaming direction, respectively and (2) an one-dimensional plot of the distribution of the particles speed in the lab frame. 

\begin{figure}[ht]
  \rotatebox{270}{ \includegraphics[width=0.62\textwidth]{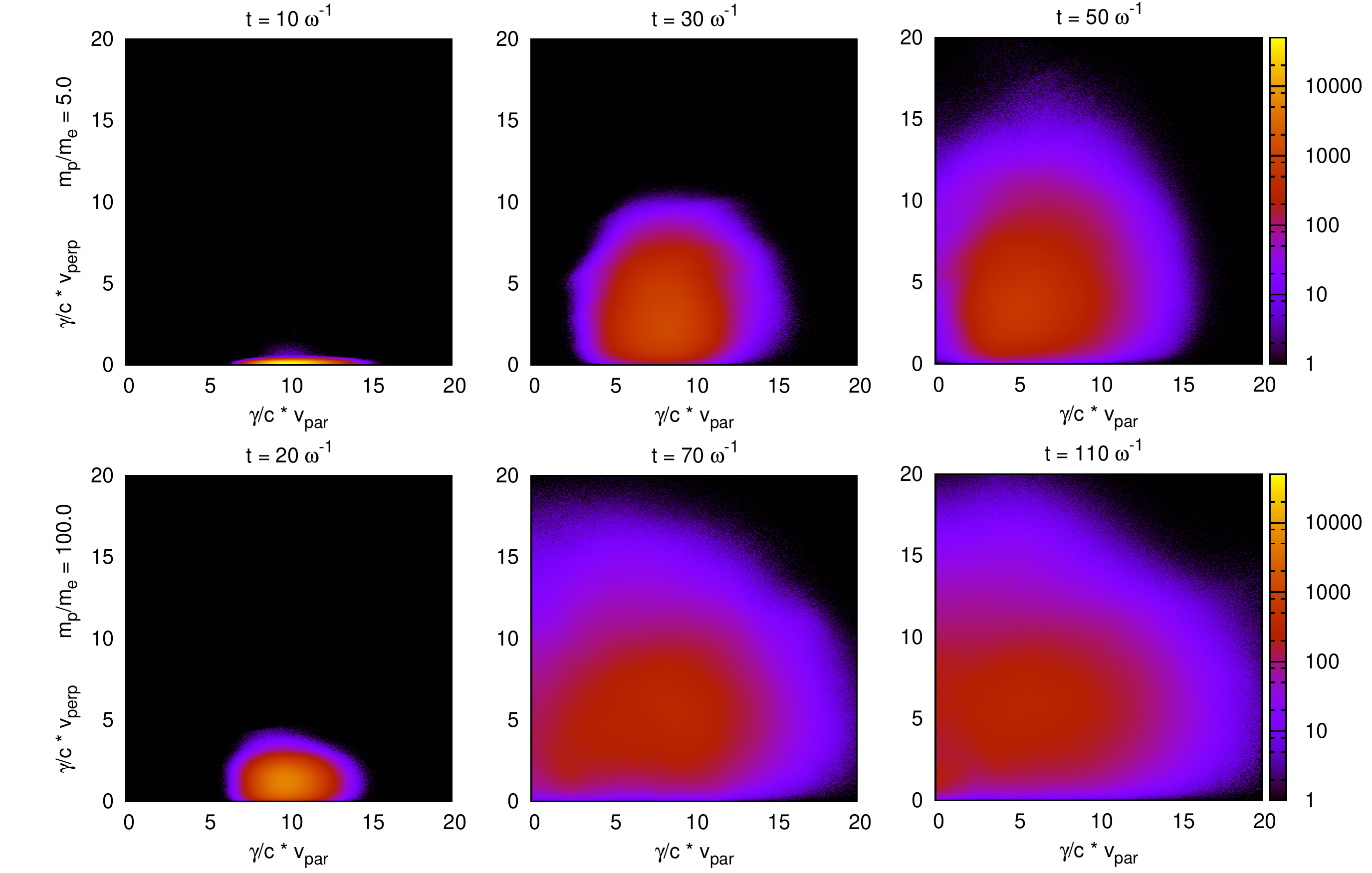} }
  \caption{\label{2Dhist_elec_pos}Development of the absolute value of the momentum distributions ($v_{\perp}/c \cdot \gamma$ over $v_{||}/c \cdot \gamma$) of all the electrons and positrons parallel and perpendicular to the direction of the original flow in the lab rest frame. First one can state that most of the particle acceleration is happening in the transverse direction. Secondly, for a mass-ratio of $m_p/m_e = 5.0$ electrons and positrons are accelerated until the simulation stops, while for the higher mass-ratio ($m_p/m_e = 100.0$) the electrons stop gaining energy around about $t = 70 \omega^{-1}$.}
\end{figure}

\begin{figure}[ht]
  \rotatebox{270}{ \includegraphics[width=0.62\textwidth]{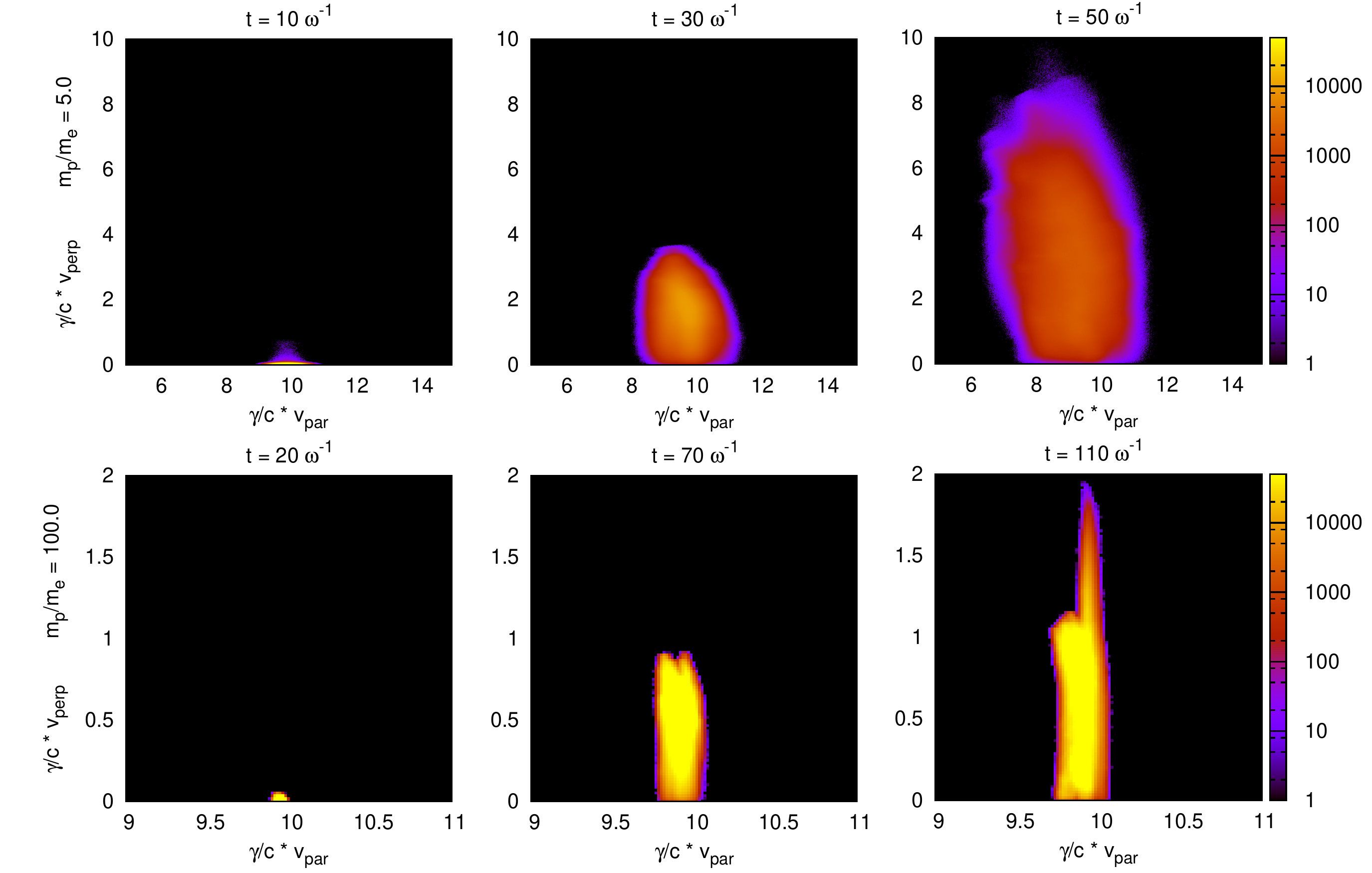} }
  \caption{\label{2Dhist_prot}Development of the absolute value of the momentum distributions ($v_{\perp}/c \cdot \gamma$ over $v_{||}/c \cdot \gamma$) of all the protons parallel and perpendicular to the direction of the original flow in the lab rest frame. First one can state that most of the particle acceleration is happening in the transverse direction. Secondly, for a mass-ratio of $m_p/m_e = 5.0$ the protons are getting accelerated over the entire timespan, while for the higher mass-ratio ($m_p/m_e = 100.0$) their acceleration starts at a later time during the simulation.}
\end{figure}

In Fig.~\ref{2Dhist_elec_pos} we show the time evolution of the electron and positron distribution (all electrons and positrons in both streams are plotted in a 2D histogram for the mass-ratios 5 (upper panels) and 100 (lower panels). The same conjuncture is illustrated in Fig.~\ref{2Dhist_prot} for the protons, but the axis are different.

As expected, the particle distribution in the early stage of the simulation (before the onset of the first instability hump) is centered at the initial Lorentz boost $\gamma = 10$ with a thermal width of $0.1 c$ for the electrons/positrons and $0.1 c \cdot (m_p/m_e)^{-1}$ for the protons. Several electrons have already been accelerated and are streaming towards higher transverse velocities. At later times in the low mass-ratio simulation one can see that both electrons and protons are getting accelerated simultaneously, as illustrated in the next two images in the upper rows of Figs. \ref{2Dhist_elec_pos} and \ref{2Dhist_prot}. As emphasized before there is no clear distinction between a low and high mass instability in this case, thus the simultaneous acceleration is clearly in agreement with the energy evolution (see Fig.~\ref{B_vergleich}).

When going to higher mass-ratios and early times (lower leftmost image of both figures) only the electrons are visible, as the protons still remain at their initial momentum. When the initial phase of the instability is over ($t \approx 70 \omega_p^{-1}$) the electrons are no longer significantly accelerated. The protons are still gaining more energy until about $t = 110 \omega_p^{-1}$, which is roughly at the maximum of the proton instability (see Fig.~\ref{B_vergleich} for comparison). This supports the idea of two almost separated instabilities.

In both cases the particle distributions show a diffusion-like behaviour: The initial distribution in parallel direction is stretched from $\gamma \approx 5$ to $\gamma \approx 15$. The particle energy is converted into perpendicular field energy which in turn accelerates the particles. We want to stress an interesting feature of the different particle distributions in the diverse simulations: Both simulations have a proton distribution which is strongly elongated in the perpendicular direction and an electron distribution whose center is below the initial $\gamma=10$ and extends more or less equal in the parallel and perpendicular direction. In the high mass case it is obvious that the electrons are partially decelerated to lower energies. 

\begin{figure}[ht]
  \includegraphics[width=0.99\textwidth]{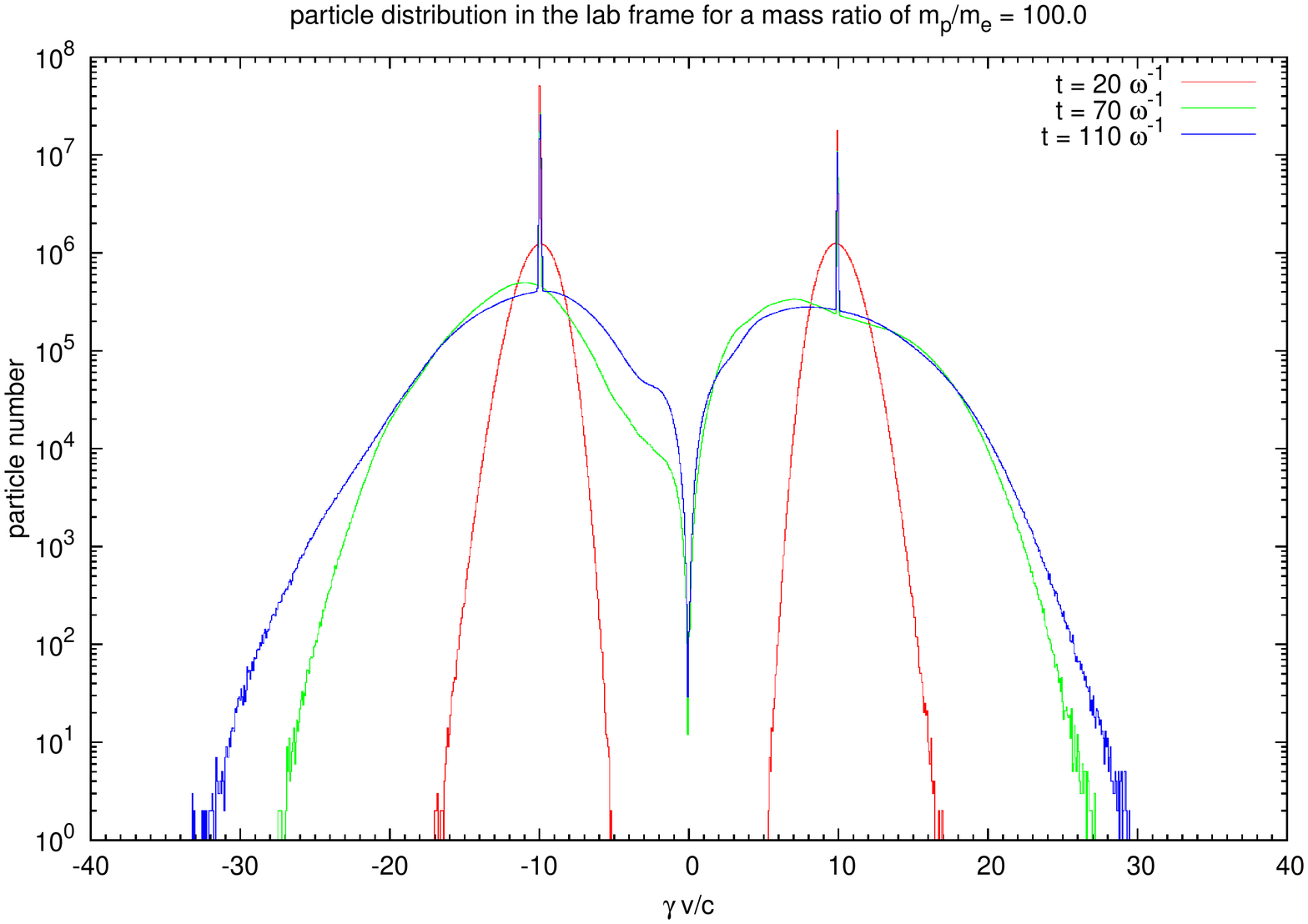}
  \caption{\label{ohneboost}Time evolution of all the particles (both proton and pair-proton stream particles) momentum distribution in the lab frame for a mass-ratio of $m_p/m_e = 100.0$. Particles moving to the left (proton stream) are shown with a negative total momentum, particles moving to the right (pair-proton stream) with a positive one. The narrow peak for each time shows the protons, the broad ones represents the electrons/positrons.}
\end{figure}

In Fig.~\ref{ohneboost} we show a one-dimensional plot of the temporal evolution of the total momentum distribution of all the particles (from both proton and pair-proton stream) in the lab frame. Particles moving in negative (positive) z-direction are plotted with a negative (positive) total momentum and the narrow peak for each time shows the protons, the broad ones represent the electrons/positrons. One can see that most of the acceleration of the electrons and positrons is happening between $20 \omega_p^{-1}$ and $70 \omega_p^{-1}$, the protons still gain more energy until about $110 \omega_p^{-1}$. This corresponds with the behaviour of the particles seen in Fig.~\ref{2Dhist_elec_pos} and in Fig.~\ref{2Dhist_prot}.

\section{\label{discussion}Discussion and summary}

We have conducted several simulations of counterstreaming plasmas with
different mass-ratios $1.0 < m_p/m_e < 100.0$ in order to investigate
the influence of the mass-ratio on the development of twostream
instabilities. We can draw two major conclusions from our simulations:
(1) the physics of acceleration in mixed counterstreaming
plasmas can be understood by using particle-in-cell simulations and (2) we are able to show that there
is indeed a strong implication of the mass-ratio on the results but in order to extrapolate the behaviour
to the physical mass-ratio one still needs to perform some simulations with higher mass-ratio of 200 or 500.

We are able to demonstrate that the mass-ratio has a qualitative and not only
quantitative effect on the simulated physics. For very low mass-ratios
only one instability develops which changes for mass-ratios
around 50. Taking these findings as a starting point we conclude that further simulations
of counterstreaming plasma have to be conducted with mass-ratios of
100 and above to find the correct physical behaviour. A quantitative
change has also been found: For mass-ratios $> 20$ the instability is
developing more slowly but also lasts longer (cf. Fig.~\ref{B_vergleich}) because the flux tubes can be sustained for a
greater timespan. This can be explained by a two stage instability,
that can be seen especially in the highest mass-ratios. Here in a
initial phase a pair instability develops as in pair-only plasmas
resulting in several flux tubes. In some of the flux tubes the (lighter)
electrons and positrons are streaming and these ones are developing much
stronger. After reaching its maximum the light-particle instability is
decreasing and some other flux tubes (carrying mostly the heavier
protons) are growing stronger (see Fig.~\ref{9_slices_100}).

Besides the investigation of the nature of the instability itself the acceleration of particles was a
subject of research. From two-dimensional plots of particle
momentum parallel versus perpendicular to the original streaming direction in the lab rest frame we concluded that most of
the particle acceleration happens in the transverse direction.
Furthermore, the two stage process can also be observed here: While
light and heavy particles are accelerated almost simultaneously for
the low mass-ratio simulations, the electrons are accelerated stronger and earlier compared to the protons in the high mass-ratio simulations. Additionally, at a later stage the
electron distribution stays almost the same while the much heavier protons still gain
energy (see Fig.s \ref{2Dhist_elec_pos} and \ref{2Dhist_prot}). This gives important
insight into the fundamental acceleration mechanism: Only particles involved in
the instability may be accelerated.

Since the instability itself and the restrictions on
numerical parameters (i.e.~the mass-ratio) are now better understood, it is 
necessary to conceive the implications on the underlying physics of astrophysical jet phenomena. From this study it is
yet not possible to impose strong limits on the possible
composition of jets, since the major difference between pair and mixed
plasma seems to be the timescale of how the instability develops.  Obviously an
extensive parameter study is necessary to cover the full range of
possible physics. Therefore in a next stage the influence of the
composition of the background and jet plasma on the particle
acceleration shall be investigated.

\acknowledgments
 TB thanks the Deutsches Zentrum f\"ur Luft- und Raumfahrt and the LISA-Germany collaboration for funding.
 OE and UG are grateful for funding from the Elite Network of Bavaria and FS would like to thank the Deutsche Forschungsgemeinschaft (DFG SP 1124/1). Simulations have been conducted at Hochleistungsrechenzentrum Stuttgart under grant ``iotmrofi''.

\bibliographystyle{plainnat}
\bibliography{paper}

\begin{thebibliography}{24}
\providecommand{\natexlab}[1]{#1}
\providecommand{\url}[1]{\texttt{#1}}
\expandafter\ifx\csname urlstyle\endcsname\relax
  \providecommand{\doi}[1]{doi: #1}\else
  \providecommand{\doi}{doi: \begingroup \urlstyle{rm}\Url}\fi

\bibitem[{Boris}(1970)]{Boris1970}
J.~P. {Boris}.
\newblock {The Acceleration Calculation from a Scalar Potential}.
\newblock \emph{Plasma Physics Laboratory Report MATT-769}, 1970.

\bibitem[{Buneman}(1958)]{1958PhRvL...1....8B}
O.~{Buneman}.
\newblock {Instability, Turbulence, and Conductivity in Current-Carrying
  Plasma}.
\newblock \emph{Physical Review Letters}, 1:\penalty0 8--9, July 1958.
\newblock \doi{10.1103/PhysRevLett.1.8}.

\bibitem[{Chang} et~al.(2008){Chang}, {Spitkovsky}, and
  {Arons}]{2008ApJ...674..378C}
P.~{Chang}, A.~{Spitkovsky}, and J.~{Arons}.
\newblock {Long-Term Evolution of Magnetic Turbulence in Relativistic
  Collisionless Shocks: Electron-Positron Plasmas}.
\newblock \emph{\apj}, 674:\penalty0 378--387, February 2008.
\newblock \doi{10.1086/524764}.

\bibitem[{Dieckmann} et~al.(2008){Dieckmann}, {Shukla}, and
  {Drury}]{2008ApJ...675..586D}
M.~E. {Dieckmann}, P.~K. {Shukla}, and L.~O.~C. {Drury}.
\newblock {The Formation of a Relativistic Partially Electromagnetic Planar
  Plasma Shock}.
\newblock \emph{\apj}, 675:\penalty0 586--595, March 2008.
\newblock \doi{10.1086/525516}.

\bibitem[{Esirkepov}(2001)]{2001CoPhC.135..144E}
T.~Z. {Esirkepov}.
\newblock {Exact charge conservation scheme for Particle-in-Cell simulation
  with an arbitrary form-factor}.
\newblock \emph{Computer Physics Communications}, 135:\penalty0 144--153, April
  2001.
\newblock \doi{10.1016/S0010-4655(00)00228-9}.

\bibitem[{Fermi}(1949)]{1949PhRv...75.1169F}
E.~{Fermi}.
\newblock {On the Origin of the Cosmic Radiation}.
\newblock \emph{Physical Review}, 75:\penalty0 1169--1174, April 1949.
\newblock \doi{10.1103/PhysRev.75.1169}.

\bibitem[{Fonseca} et~al.(2008){Fonseca}, {Martins}, {Silva}, {Tonge}, {Tsung},
  and {Mori}]{2008PPCF...50l4034F}
R.~A. {Fonseca}, S.~F. {Martins}, L.~O. {Silva}, J.~W. {Tonge}, F.~S. {Tsung},
  and W.~B. {Mori}.
\newblock {One-to-one direct modeling of experiments and astrophysical
  scenarios: pushing the envelope on kinetic plasma simulations}.
\newblock \emph{Plasma Physics and Controlled Fusion}, 50\penalty0
  (12):\penalty0 124034--+, December 2008.
\newblock \doi{10.1088/0741-3335/50/12/124034}.

\bibitem[{Frederiksen} et~al.(2004){Frederiksen}, {Hededal}, {Haugb{\o}lle},
  and {Nordlund}]{2004ApJ...608L..13F}
J.~T. {Frederiksen}, C.~B. {Hededal}, T.~{Haugb{\o}lle}, and {\AA}.~{Nordlund}.
\newblock {Magnetic Field Generation in Collisionless Shocks: Pattern Growth
  and Transport}.
\newblock \emph{\apjl}, 608:\penalty0 L13--L16, June 2004.
\newblock \doi{10.1086/421262}.

\bibitem[{Hededal} and {Nishikawa}(2005)]{2005ApJ...623L..89H}
C.~B. {Hededal} and K.-I. {Nishikawa}.
\newblock {The Influence of an Ambient Magnetic Field on Relativistic
  collisionless Plasma Shocks}.
\newblock \emph{\apjl}, 623:\penalty0 L89--L92, April 2005.
\newblock \doi{10.1086/430253}.

\bibitem[{Hededal} et~al.(2004){Hededal}, {Haugb{\o}lle}, {Frederiksen}, and
  {Nordlund}]{2004ApJ...617L.107H}
C.~B. {Hededal}, T.~{Haugb{\o}lle}, J.~T. {Frederiksen}, and {\AA}.~{Nordlund}.
\newblock {Non-Fermi Power-Law Acceleration in Astrophysical Plasma Shocks}.
\newblock \emph{\apjl}, 617:\penalty0 L107--L110, December 2004.
\newblock \doi{10.1086/427387}.

\bibitem[{Hockney} and {Eastwood}(1988)]{1988csup.book.....H}
R.~W. {Hockney} and J.~W. {Eastwood}.
\newblock \emph{{Computer simulation using particles}}.
\newblock 1988.

\bibitem[{Martins} et~al.(2009){Martins}, {Fonseca}, {Silva}, and
  {Mori}]{2009ApJ...695L.189M}
S.~F. {Martins}, R.~A. {Fonseca}, L.~O. {Silva}, and W.~B. {Mori}.
\newblock {Ion Dynamics and Acceleration in Relativistic Shocks}.
\newblock \emph{\apjl}, 695:\penalty0 L189--L193, April 2009.
\newblock \doi{10.1088/0004-637X/695/2/L189}.

\bibitem[{Nishikawa} et~al.(2009){Nishikawa}, {Medvedev}, {Zhang}, {Hardee},
  {Niemiec}, {Nordlund}, {Frederiksen}, {Mizuno}, {Sol}, and
  {Fishman}]{2009arXiv0901.4058N}
K.~. {Nishikawa}, M.~{Medvedev}, B.~{Zhang}, P.~{Hardee}, J.~{Niemiec},
  A.~{Nordlund}, J.~{Frederiksen}, Y.~{Mizuno}, H.~{Sol}, and G.~J. {Fishman}.
\newblock {Radiation from relativistic jets in turbulent magnetic fields}.
\newblock \emph{ArXiv e-prints}, January 2009.

\bibitem[{Nishikawa} et~al.(2003){Nishikawa}, {Hardee}, {Richardson}, {Preece},
  {Sol}, and {Fishman}]{2003ApJ...595..555N}
K.-I. {Nishikawa}, P.~{Hardee}, G.~{Richardson}, R.~{Preece}, H.~{Sol}, and
  G.~J. {Fishman}.
\newblock {Particle Acceleration in Relativistic Jets Due to Weibel
  Instability}.
\newblock \emph{\apj}, 595:\penalty0 555--563, September 2003.
\newblock \doi{10.1086/377260}.

\bibitem[{Nishikawa} et~al.(2005){Nishikawa}, {Hardee}, {Richardson}, {Preece},
  {Sol}, and {Fishman}]{2005ApJ...622..927N}
K.-I. {Nishikawa}, P.~{Hardee}, G.~{Richardson}, R.~{Preece}, H.~{Sol}, and
  G.~J. {Fishman}.
\newblock {Particle Acceleration and Magnetic Field Generation in
  Electron-Positron Relativistic Shocks}.
\newblock \emph{\apj}, 622:\penalty0 927--937, April 2005.
\newblock \doi{10.1086/428394}.

\bibitem[{Nishikawa} et~al.(2006){Nishikawa}, {Hardee}, {Hededal}, and
  {Fishman}]{2006ApJ...642.1267N}
K.-I. {Nishikawa}, P.~E. {Hardee}, C.~B. {Hededal}, and G.~J. {Fishman}.
\newblock {Acceleration Mechanics in Relativistic Shocks by the Weibel
  Instability}.
\newblock \emph{\apj}, 642:\penalty0 1267--1274, May 2006.
\newblock \doi{10.1086/501426}.

\bibitem[{Nishikawa} et~al.(2008){Nishikawa}, {Niemiec}, {Sol}, {Medvedev},
  {Zhang}, {Nordlund}, {Frederiksen}, {Hardee}, {Mizuno}, {Hartmann}, and
  {Fishman}]{2008AIPC.1085..589N}
K.-I. {Nishikawa}, J.~{Niemiec}, H.~{Sol}, M.~{Medvedev}, B.~{Zhang},
  {\AA}.~{Nordlund}, J.~{Frederiksen}, P.~{Hardee}, Y.~{Mizuno}, D.~H.
  {Hartmann}, and G.~J. {Fishman}.
\newblock {New Relativistic Particle-In-Cell Simulation Studies of Prompt and
  Early Afterglows from GRBs}.
\newblock In F.~A. {Aharonian}, W.~{Hofmann}, and F.~{Rieger}, editors,
  \emph{American Institute of Physics Conference Series}, volume 1085 of
  \emph{American Institute of Physics Conference Series}, pages 589--593,
  December 2008.
\newblock \doi{10.1063/1.3076741}.

\bibitem[{Ramirez-Ruiz} et~al.(2007){Ramirez-Ruiz}, {Nishikawa}, and
  {Hededal}]{2007ApJ...671.1877R}
E.~{Ramirez-Ruiz}, {K.-I.} {Nishikawa}, and C.~B. {Hededal}.
\newblock {$e^{+/-}$ Pair Loading and the Origin of the Upstream Magnetic Field
  in GRB Shocks}.
\newblock \emph{\apj}, 671:\penalty0 1877--1885, December 2007.
\newblock \doi{10.1086/522072}.

\bibitem[{Silva} et~al.(2003){Silva}, {Fonseca}, {Tonge}, {Dawson}, {Mori}, and
  {Medvedev}]{2003ApJ...596L.121S}
L.~O. {Silva}, R.~A. {Fonseca}, J.~W. {Tonge}, J.~M. {Dawson}, W.~B. {Mori},
  and M.~V. {Medvedev}.
\newblock {Interpenetrating Plasma Shells: Near-equipartition Magnetic Field
  Generation and Nonthermal Particle Acceleration}.
\newblock \emph{\apjl}, 596:\penalty0 L121--L124, October 2003.
\newblock \doi{10.1086/379156}.

\bibitem[{Spitkovsky}(2008{\natexlab{a}})]{2008ApJ...673L..39S}
A.~{Spitkovsky}.
\newblock {On the Structure of Relativistic Collisionless Shocks in
  Electron-Ion Plasmas}.
\newblock \emph{\apjl}, 673:\penalty0 L39--L42, January 2008{\natexlab{a}}.
\newblock \doi{10.1086/527374}.

\bibitem[{Spitkovsky}(2008{\natexlab{b}})]{2008ApJ...682L...5S}
A.~{Spitkovsky}.
\newblock {Particle Acceleration in Relativistic Collisionless Shocks: Fermi
  Process at Last?}
\newblock \emph{\apjl}, 682:\penalty0 L5--L8, July 2008{\natexlab{b}}.
\newblock \doi{10.1086/590248}.

\bibitem[{Taflove} and {Hagness}(2005)]{Taflove}
A.~{Taflove} and S.~C. {Hagness}.
\newblock \emph{{Computational Electrodynamics: The Finite-Difference
  Time-Domain Method, 3rd ed.}}
\newblock 2005.

\bibitem[{Weibel}(1959)]{1959PhRvL...2...83W}
E.~S. {Weibel}.
\newblock {Spontaneously Growing Transverse Waves in a Plasma Due to an
  Anisotropic Velocity Distribution}.
\newblock \emph{Physical Review Letters}, 2:\penalty0 83--84, February 1959.
\newblock \doi{10.1103/PhysRevLett.2.83}.

\bibitem[{Yee}(1966)]{1966ITAP...14..302Y}
K.~{Yee}.
\newblock {Numerical solution of inital boundary value problems involving
  maxwell's equations in isotropic media}.
\newblock \emph{IEEE Transactions on Antennas and Propagation}, 14:\penalty0
  302--307, May 1966.
\newblock \doi{10.1109/TAP.1966.1138693}.

\end{thebibliography}

\end{document}